\documentclass[twocolumn,amsmath,amssymb,pre,superscriptaddress]{revtex4}
\usepackage{graphicx}
\usepackage{dcolumn}
\usepackage{bm}
\usepackage{subfigure}
\usepackage{booktabs}
\usepackage{url}
\usepackage{color}
\newcounter{one}
\newcommand{\inter}{\curvearrowright}
\newcommand{\intra}{\circlearrowright}

\newcommand{\affA}{Department of Computational Intelligence and Systems Science, 
Tokyo Institute of Technology, 4259-G5-22, Nagatsuta-cho, Midori-ku, Yokohama, Kanagawa, 226-8502, Japan}
\newcommand{\affB}{Integrated Science Lab, Department of Physics, Ume{\aa} University, SE-901 87 Ume{\aa}, Sweden}

\begin{document}







\title{Estimating the resolution limit of the map equation in community detection}

\author{Tatsuro Kawamoto}
\affiliation{\affA}
\author{Martin Rosvall}
\affiliation{\affB}
\date{\today}

\begin{abstract}
A community detection algorithm is considered to have a resolution limit if the scale of the smallest modules that can be resolved depends on the size of the analyzed subnetwork. The resolution limit is known to prevent some community detection algorithms from accurately identifying the modular structure of a network. In fact, any global objective function for measuring the quality of a two-level assignment of nodes into modules must have some sort of resolution limit or an external resolution parameter. However, it is yet unknown how the resolution limit affects the so-called map equation, which is known to be an efficient objective function for community detection. We derive an analytical estimate and conclude that the resolution limit of the map equation is set by the total number of links between modules instead of the total number of links in the full network as for modularity. This mechanism makes the resolution limit much less restrictive for the map equation than for modularity; in practice, it is orders of magnitudes smaller.
Furthermore, we argue that the effect of the resolution limit often results from shoehorning multilevel modular structures into two-level descriptions. As we show, the hierarchical map equation effectively eliminates the resolution limit for networks with nested multilevel modular structures. 
\end{abstract}
 
\maketitle

\section{Introduction}
The ability to detect the community structure of networks plays an important role in the analysis of complex systems.
Therefore, researchers have developed a suite of community detection algorithms based on different principles or heuristics \cite{Newman2004,Palla2005,Rosvall2008,Delvenne2010,Mucha2010,Luxburg2007,Fortunato201075}. 
While graph partitioning methods require the number of modules as input, 
community detection methods, such as \textit{modularity} \cite{Newman2004} and the \textit{map equation} \cite{Rosvall2008}, intrinsically identify the number of modules \cite{Fortunato201075}. 
The widely used method of modularity has been studied extensively \cite{Newman2006PRE,Newman2006PNAS,Blondel2008,Delvenne2010,Schaub2012}, but less is known about the inner workings of the flow-based and information-theoretic map equation \cite{Schaub2012a,Lambiotte2012}, despite its strong performance on benchmark networks with densely connected components when companioned with its search algorithm Infomap \cite{Lancichinetti2009,Aldecoa2013}. Interestingly, the resolution limit \cite{Fortunato2007}, which causes small modules to aggregate in larger modules in modularity maximization and can lead to poor performance in resolving actual communities of real networks, seems to have an unnoticeable effect on the map equation \cite{lancichinetti2011limits}. Since any global objective function for two-level community detection  must have a resolution limit \cite{traag2013significant}, or an external resolution parameter \cite{ronhovde2010local,traag2011narrow}, it is important to understand how the map equation succeeds at suppressing the effect of the resolution limit.
In this paper, we analytically derive the resolution limit of the map equation and show why the map equation can resolve a much wider range of module sizes than modularity can. 
Although the community structure does not necessarily mean the densely connected components, because the resolution limit is about a detectability of densely connected components, we focus on such a case. 

The resolution limit is the consequential downside of methods that intrinsically identify a resolution scale in a network to determine the number of modules.
We conceptually illustrate this fact with the two-level map equation applied to the global road network. The map equation framework seeks an optimal modular description of a random walker on the network. This maximum compression is achieved by balancing the description length of movements within and between modules. With many small modules, representing city neighborhoods, for example, the within-module description length is short at the cost of a very long between-module description length. Contrarily, with few large modules, representing continents, for example, the between-module description length is short at the cost of a very long within-module description length. Consequently, the optimal two-level description length is achieved by identifying modules of intermediate sizes, such as neighborhoods aggregated into cities. If, however, only a subnetwork was analyzed, such as the network of a single city, modules would likely correspond to neighborhoods. 
In contrast, with a so-called resolution limit-free method, and given an external and fixed resolution parameter, 
the modules identified in the subnetwork could also be identified in the full network \cite{ronhovde2010local,traag2011narrow}.
However, circumventing the resolution limit does, of course, not in itself imply good performance in resolving actual communities of real networks \cite{lancichinetti2011limits}. Moreover, for real networks the network itself must be used to set the resolution parameter, and the method is no longer resolution limit-free \cite{traag2013significant}. This example makes clear that no two-level community detection method is resolution limit-free in practice and that the resolution limit can arise because a two-level method is applied to a multilevel structure with nested modules. We argue that this case should be considered unproblematic and show later that the natural solution is to use a multilevel community detection method \cite{Rosvall2011}.

However, it is more problematic when a method aggregates small modules in a plain modular structure. In this case, a mechanistic understanding of how a method performs is critical for successful application \cite{Fortunato2007,Kumpula2007,Good2010,Berry2011}. For example, in a network with $L$ links, Fortunato and Barth\'{e}lemy \cite{Fortunato2007} showed that modularity may fail to detect a module of size less than about $\sqrt{L}$ links. This limit is a result of the intrinsic scale of the method \cite{Schaub2012,Schaub2012a}. Similarly, for a stochastic block model with model selection based on the minimum description principle, the corresponding block size was found to be about $\sqrt{N}$ nodes \cite{Peixoto2013}.  It has been experimentally verified that the map equation has an upper limit of detectability; Infomap splits non-clique structures with large diameters, such as strings and lattices \cite{Schaub2012a}, but the resolution limit at which modules cannot be fully resolved by the map equation is still unknown.
Unlike modularity and stochastic block models, which build on configuration and generative models, respectively, the map equation operates by compressing a modular description of flow on the network. 
The different machinery makes it especially interesting to better understand the effect of the resolution limit on the map equation.

\section{The map equation}
To derive the limit and show how it depends on the network structure, we begin by reviewing the machinery of the map equation, which takes advantage of the fundamental duality in information theory between finding regularities in data and compressing the data \cite{shannon1948mathematical}. For community detection, the regularities naturally correspond to the modules in a network. 

The purest form of a modular network consists of isolated cliques of nodes. With a random walker as a proxy for dynamics on a network, its movements to any node from any other node in the clique occur with equal probability. That is, node visits are independent and identically distributed. Consequently, the map equation's underlying code structure is based precisely on the assumption of independent and identically distributed node visits and module entries and exits such that it can efficiently compress the description length of the random walker's trajectory in a modular network. 

Specifically, given a partition $\mathsf M$ of nodes $i$ assigned to modules $\boldsymbol{\imath}=1,2,\ldots,m$, the map equation measures the per-step average description length $L({\mathsf M})$ of dynamics on a network. 
The movements are encoded as follows: 
$m$ \emph{module codebooks}, one for each module, map node visits within modules and exits from modules to codewords for describing movements within modules, and one \emph{index codebook} maps entries into modules to codewords for describing movements between modules. The length of the codewords are optimally derived from the rates of the corresponding movements they describe. 
However, the explicit description with codewords is not necessary for taking advantage of the duality between finding regularities in data and compressing the data. Instead, only the description length is required. Therefore, for a given partition of the network, the map equation simply measures the average codelength of each codebook and weights them by how often they are used.
In any case, the modular partition that provides the most efficient compression of the random walker's movements also best captures the community structure with respect to the dynamics on the network.

According to Shannon's source coding theorem \cite{shannon1948mathematical}, the average minimum description length of each codebook directly from the associated probability distribution $X$ of the corresponding node-visit and module-transition rates in terms of the Shannon entropy $H(X)=-\sum_i{P(x_i)}\log_2{P(x_i)}$.
Then, the complete average description length $L(\mathsf{M})$ is simply the sum of the average description length of all codebooks weighted by their rate of use. That is, 
\begin{align}
L(\mathsf{M}) = q_{\curvearrowleft} H(\mathcal{Q}) + \sum_{i=1}^{m}p_{i\intra}H(\mathcal{P}_i). \label{FundamentalL1}
\end{align} 
The first term is the average description length of the index codebook. Its rate of use $q_{\curvearrowleft} = \sum_{i=1}^{m} q_{i\curvearrowleft}$ is the sum of the  entering rates $q_{i\curvearrowleft}$ into each module $i$, and 
$H(\mathcal{Q}) = - \sum_{i=1}^{m} ( q_{i\curvearrowleft} / q_{\curvearrowleft}) \log( q_{i\curvearrowleft} / q_{\curvearrowleft})$ is the average description length of the index codebook given by the Shannon entropy of the entering rates into the modules. 
We use $\log$ base $2$ throughout this paper. 
The second term is the average description length of the module codebooks. 
The rate of use of module codebook $i$, $p_{i\intra} = q_{i\inter} + \sum_{\alpha \in i} p_{\alpha}$, is the sum of the exiting rate $q_{i\inter}$ and the  visiting rates $p_{\alpha}$ of all nodes $\alpha$ in module $i$, and
$H(\mathcal{P}^{i}) = -( q_{i\inter} / p_{i\intra} ) \log( q_{i\inter} / p_{i\intra} ) 
 -\sum_{\alpha \in i} (p_{\alpha} / p_{i\intra}) \log(p_{\alpha} / p_{i\intra})$ 
 is the average description length of module codebook $i$. All rates are evaluated at stationarity, such that for undirected networks the rate of entering $q_{i\curvearrowleft}$ and exiting $q_{i\inter}$ a given module $i$ are the same. Using this equality and expansion of Eq.~(\ref{FundamentalL1}) gives  
\begin{align}
L({\mathsf M}) 
 &= q_{\inter} \log q_{\inter} 
 -2 \sum_{i=1}^{m} q_{i\inter} \log q_{i\inter} \nonumber\\
& +\sum_{i=1}^{m} p_{i\intra} \log p_{i\intra} -\sum_{\alpha} p_{\alpha} \log p_{\alpha}. \label{QualityFunction}
\end{align}
Since the sum in the last term runs over all nodes, it does not depend on the choice of partition. 

For undirected, unweighted networks, the exit probability $q_{i\inter}$ from module $i$ and the rate of use $p_{i\intra}$ of module codebook $i$ 
can be expressed in terms of the number of links as 
\begin{align}
p_{\alpha} &= \frac{k_{\alpha}}{K}, \label{pa}\\
q_{i\inter} &= \sum_{\beta \notin i} \sum_{\alpha \in i} T_{\beta \alpha} p_{\alpha} = \frac{l^{\mathrm{out}}_{i}}{K}, \label{qinter}\\
p_{i\intra} &= q_{i\inter} + \sum_{\alpha \in i} p_{\alpha} = 2\frac{l_{i} + l^{\mathrm{out}}_{i} }{K}, \label{pi}
\end{align}
where $k_{\alpha}$ is the degree of node $\alpha$, $T_{\beta \alpha}$ is the transition probability that the random walker moves from node $\alpha$ to node $\beta$,
$l^{\mathrm{out}}_{i}$ is the number of links which connect the nodes in module $i$ with nodes in other modules, 
$l_{i}$ is the number of links within module $i$, 
and $K = 2L=\sum_{i}(2l_{i}+l^{\mathrm{out}}_{i})$ is the total degree of the network. 
Substituting these expressions into Eq.~(\ref{QualityFunction}) gives
\begin{align}
&L({\mathsf M}) 
= \frac{1}{K} \left[ 2 C \log 2 C 
-2 \sum^{m}_{i=1} l^{\mathrm{out}}_{i} \log l^{\mathrm{out}}_{i} + K + 2C \right. \nonumber\\
&\hspace{10pt} \left. +2 \sum^{m}_{i=1} (l_{i} + l^{\mathrm{out}}_{i}) \log (l_{i} + l^{\mathrm{out}}_{i} )
-\sum_{\alpha} k_{\alpha} \log k_{\alpha}
\right], \label{generalL}
\end{align}
where $k_{\alpha}$ is the degree of the node $\alpha$ and $C$ is the \textit{cut size} \cite{Fortunato201075,Luxburg2007}, i.e., $2 C = \sum^{m}_{i=1} l^{\mathrm{out}}_{i}$. 

\section{The resolution limit of the map equation}
The resolution limit of the map equation depends on which partitions are favored when the map equation is minimized. 
In general, to minimize Eq.~(\ref{generalL}), we update the partition ${\mathsf M} = {\mathsf A}$ to a nearby partition ${\mathsf M} = {\mathsf B}$ if $\Delta L({\mathsf M}) = L({\mathsf B}) - L({\mathsf A}) < 0$. 
From Eq.~(\ref{generalL}), we readily see that there is no resolution limit caused by the total degree $K$, because it never changes the sign of $\Delta L({\mathsf M})$. 
Instead, the resolution limit must depend on the cut size, which controls the only global term.
In other words, as long as the update conserves the cut size $C$, 
there is no restriction from the global structure under an arbitrary update.
Note that the resolution limit of the stochastic block model with model selection based on the minimum description principle depends on the number of nodes in the network \cite{Peixoto2013}.
Therefore, the dependence on the cut size is not a result of the information-theoretic nature of the map equation per se, but rather its machinery to describe flow trajectories on networks.

To identify which partition updates will be accepted, we denote an element in the two sums over modules in Eq.~(\ref{generalL}) by 
\begin{align}
\mathcal{L}_{i} = - l^{\mathrm{out}}_{i} \log l^{\mathrm{out}}_{i} + (l_{i} + l^{\mathrm{out}}_{i}) \log (l_{i} + l^{\mathrm{out}}_{i} ), \label{localL}
\end{align}
and denote the total change over all modules before and after the update by
\begin{align}
R = \sum_{i^{\prime}} \mathcal{L}_{i^{\prime}}({\mathsf B}) - \sum_{i} \mathcal{L}_{i}({\mathsf A}).
\label{Rdef}
\end{align}
We now consider a local update where the cut size is decreased by a small $\delta$, such that $C \gg \delta > 0$. 
Then
\begin{align}
\Delta L({\mathsf M}) 
&= \frac{1}{K} [ 
2(C - \delta) \log \left( 2(C - \delta) \right) - 2C \log \left( 2C \right) \nonumber\\
&\hspace{140pt} +2 R - 2\delta ] \label{generalDeltaLexact}\\
&\simeq \frac{2}{K} \left[ 
- \delta \left( 2 + \log \left( \mathrm{e} \, C \right) \right) + R \right], 
\label{generalDeltaL}
\end{align}
where $\mathrm{e}$ is the basis of natural logarithm. 
Therefore, any local update should be accepted if 
\begin{align}
R \lesssim 
\delta \left( 2 + \log \left(\mathrm{e} \, C \right) \right). 
\label{GeneralResolutionLimit}
\end{align}
For updates that increase the cut size, i.e., $\delta < 0$, on the other hand, 
no local update will be accepted as long as the cut size $C$ is sufficiently large. 

\begin{figure}[!tbp]
\centering
\includegraphics[width=0.4\columnwidth]{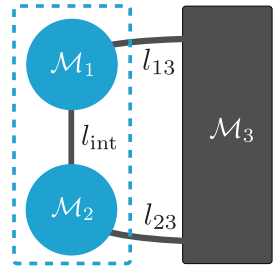}
\caption{
(Color online) A schematic picture for the greedy update of two modules $\mathcal{M}_{1}$ and $\mathcal{M}_{2}$. 
Note that $\mathcal{M}_{3}$ may consist of many modules. 
}
\label{GeneralGreedy}
\end{figure}
To identify the resolution limit, we parallel the analysis of Ref.~\cite{Fortunato2007} and 
pinpoint a partition at the point where the map equation can resolve small modules.
As shown in Fig.~\ref{GeneralGreedy}, we use a partition ${\mathsf A}$ with two modules $\mathcal{M}_{1}$ and $\mathcal{M}_{2}$ connected with 
$l_{\mathrm{int}}$ links between them, and $l_{13}$ and $l_{23}$ links, respectively, with the rest of the network $\mathcal{M}_{3}$. 
Note that $\mathcal{M}_{3}$ may consist of many modules, so that 
Fig.~\ref{GeneralGreedy} represents a completely general situation. 
In partition ${\mathsf B}$, the two modules $\mathcal{M}_{1}$ and $\mathcal{M}_{2}$ are
merged into a single module $\mathcal{M}_{12}$.
We then consider an update from partition ${\mathsf A}$
to partition ${\mathsf B}$. From Eq.~(\ref{Rdef}) we have:
\begin{align}
&R = 
- l^{\mathrm{out}}_{3} \log l^{\mathrm{out}}_{3} 
+ l^{\mathrm{out}}_{2} \log l^{\mathrm{out}}_{2} 
+ l^{\mathrm{out}}_{1} \log l^{\mathrm{out}}_{1} \nonumber\\
&\hspace{2pt} + (l_{1} + l^{\mathrm{out}}_{1} + l_{2} + l^{\mathrm{out}}_{2} - l_{\mathrm{int}} ) 
\log (l_{1} + l^{\mathrm{out}}_{1} + l_{2} + l^{\mathrm{out}}_{2} - l_{\mathrm{int}}) \nonumber\\
&\hspace{2pt} - (l_{2} + l^{\mathrm{out}}_{2}) \log (l_{2} + l^{\mathrm{out}}_{2}) 
- (l_{1} + l^{\mathrm{out}}_{1}) \log (l_{1} + l^{\mathrm{out}}_{1}), \label{GreedyR}
\end{align}
where $l^{\mathrm{out}}_{3} = l_{13} + l_{23}$. We now consider the extreme case in which $l_{\mathrm{int}} = 1$, i.e., $\delta = 1$, 
and set the sizes of two modules equal, i.e., $l_{1} = l_{2}=l_{c}$, 
because it maximizes $R$ in $0 < l_{2} \le l_{1}$ for a fixed $l_{1}$. 
We also set $l_{13} = l_{23} = h$. 
Then, using the assumption that $l_{c}+h \gg 1$, we have 
\begin{align}
R 
&\simeq 1 + 2 \left[ l_{c} + (1+h)\log(1+h) -h\log h \right] \nonumber\\
&\hspace{10pt} - \log \left[ \mathrm{e}(l_{c}+h) \right]. 
\label{GreedyRmax}
\end{align}
Assuming that $h=1$ and using Eq.~(\ref{GeneralResolutionLimit}), 
we obtain the inequality for the resolution limit 
\begin{align}
\frac{4^{l_{c}}}{l_{c}+1} \lesssim C, \label{GreedyResolutionLimit}
\end{align}
where we have dropped a small constant factor $8/\mathrm{e}^{2} \simeq 1.08$ on the left-hand side, in order to highlight the basic scaling. 
Assuming that $h \gg 1$, 
$(1+h)\log(1+h) -h\log h \simeq \log\left[ \mathrm{e}(1+h)\right]$, we instead have 
\begin{align}
\frac{(1+h)^{2} 4^{l_{c}}}{2(l_{c}+h)} \lesssim C. \label{hGreedyResolutionLimit}
\end{align}
Accordingly, the map equation fails to detect a module with less than $l_{c}$ links 
whenever the cut size $C$ satisfies the above conditions, 
provided that the module is adjacent to modules of equal or smaller size. 
Note that Eqs.~(\ref{GreedyResolutionLimit}) and (\ref{hGreedyResolutionLimit}) apply 
only when evaluated close to the global minimum of the map equation.
Otherwise, they are only practical restrictions during an optimization process. 
Furthermore, with $l_{1} = l_{2} = l$ and $l_{13} = l_{23} = h$ in Fig.~\ref{GeneralGreedy},
the following two examples are worth mentioning:
First, if $l_{\mathrm{int}} = 0$, $\Delta L({\mathsf M}) = 4 l/K > 0$, and disconnected modules never merge with other modules, as they should not.
Second, if $\mathcal{M}_{3}$ is a single module, $\Delta L(h=1) > 0$ for $l \ge 2$ and 
$\partial \Delta L / \partial h > 0$, and $\Delta L(h \rightarrow \infty) = 4(l-1)/K$ for any $l$, 
such that the map equation can detect modules of arbitrary size with $l \ge 2$. 

\begin{figure}[tbp]
\centering
\includegraphics[width=\columnwidth]{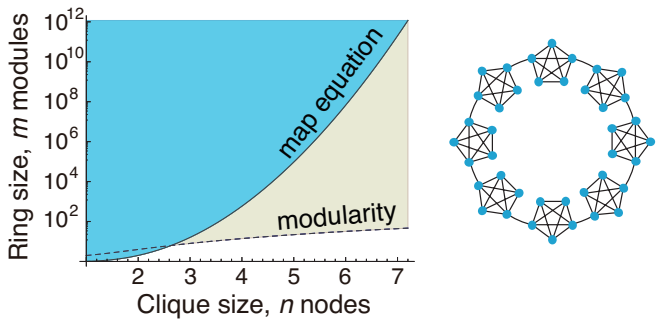}
\caption{(Color online) Detectable region of $m$ modules, each forming a clique with $n$ nodes.  
		In the inset network, $n=5$ and $m=8$. 
		The solid line is the resolution limits of the map equation according to Eq.~(\ref{GreedyResolutionLimit}), while 
		the dashed line is the resolution limit of modularity, $m = n(n-1)+2$ \cite{Fortunato2007}. 
		}
\label{RingOfCliques}
\end{figure}

To illustrate that the resolution limit of the map equation is much smaller in practice and less restrictive than it is for modularity, 
we consider a simple modular network shown in the inset of Fig.~\ref{RingOfCliques}. The network consists of a ring of $m$ modules, each forming a clique with $n$ nodes. For this network, $C = m$. 
The resolution limit of the map equation is in practice many orders of magnitudes smaller than the resolution limit of modularity.
For clique size 6, for example, the map equation can resolve modules in a ring network that is millions of times larger.
If this intrinsic scale of resolution set by using one node codeword per step is not appropriate for the problem at hand, and a multilevel solution is not an option, the scale can be modified by the Markov time, which effectively works as an intrinsic resolution parameter \cite{Schaub2012,Schaub2012a}.
More importantly, the resolution limit does not depend on the number of links in the network, as for modularity, but on the cut size. 
This feature makes the resolution limit less restrictive with important performance implications. 

\section{The resolution limit of the hierarchical map equation}
We now turn to the hierarchical map equation \cite{Rosvall2011}, the multilevel generalization of the two-level method described above.
For example, the three-level map equation can consider movements within and between supermodules, modules in supermodules, and nodes in modules. In the general case, this hierarchy of nested modules can be extended locally and independently between branches as long as it reduces the minimal description length. 
Because modification of a partition at a certain level of a branch only influences the description length of movements within and between affected modules,
$\Delta L({\mathsf M})$ for the hierarchical map equation turns out to be analogous to that of the two-level map equation (see Sec.~\ref{AppendixHierarchicalMapEquation-General} of the Appendix for details).
However, and importantly, the resolution limit now depends on the structure of the associated supermodule rather than on the structure of the full network. 
As a result, in the absence of a nested multilevel modular structure for the hierarchical map equation to capitalize on, the resolution limit of the two-level method remains. 
However, with a sufficiently pronounced nested multilevel modular structure, the hierarchical map equation will resolve all modules (see Secs.~\ref{AppendixHierarchicalMapEquation-SupermoduleGeneration} and \ref{AppendixHierarchicalMapEquation-HierarchicalRing} of the Appendix for details). 
A similar effect is true also for methods based on generative models. For example, the typical detectable block size decreases from $\sim \sqrt{N}$ to $\sim \ln N$ for a hierarchical generalization of the stochastic block model mentioned above \cite{peixoto2013hierarchical}. 
In analogy with the two-level analysis, since the hierarchical map equation determines the number of levels of each branch intrinsically based on the network structure rather than with an external parameter, it is not resolution limit-free in the sense that analysis of the full network or a subnetwork necessarily would give the same result \cite{traag2011narrow}. Nevertheless, as long as the network has a pronounced nested multilevel modular structure, modules at the finest level will be resolved.

\begin{figure}[tbp]
\centering
\includegraphics[width=0.95\columnwidth]{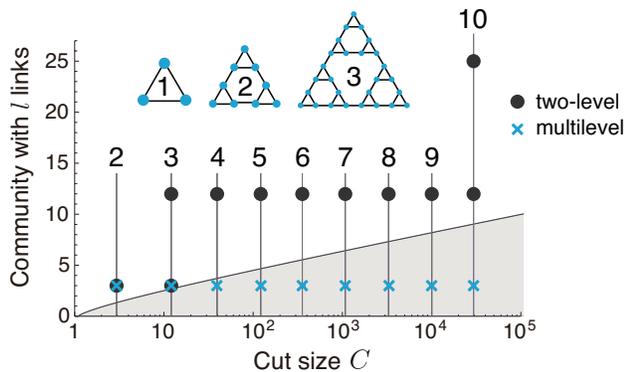}
\caption{
(Color online) Detected sizes of modules for each Sierpinski triangle of different size $l$
with the two-level method (circular points) and the multilevel method (cross points). 
For example, the two-level method detects modules of 3 links and 12 links when the hierarchy of the Sierpinski triangle is three. 
The Sierpinski triangles up to three levels are illustrated at the top. 
The boundary of the shaded region shows the resolution limit of the module size obtained with Eq.~(\ref{GreedyResolutionLimit}). 
}
\label{SierpinskiTriangles}
\end{figure}

To demonstrate the performance of the hierarchical map equation, we use Sierpinski triangles, as shown in the inset of Fig.~\ref{SierpinskiTriangles}. 
Using the code distributed at \url{http://www.mapequation.org},
we identify modules in the Sierpinski triangles of different sizes with the two-level and multilevel methods. 
For the multilevel method, we focus on the results at the finest level. 
While the result of the two-level method is harmed by the resolution limit, 
the multilevel method detects the triangles at the lowest level for any network size.
Also, cliques in a ring of cliques, as illustrated in Fig.~\ref{RingOfCliques}, are resolved for any network size (see Sec.~\ref{AppendixHierarchicalMapEquation-HierarchicalRing} of the Appendix for analytical derivation).
Furthermore, as we see in the next section, the relaxed resolution limit can be observed in real networks as well. 
Therefore, we conclude that the hierarchical map equation effectively eliminates the resolution limit for networks with nested multilevel modular structures.

\section{Effects of the resolution limit: module size distributions of real networks} \label{RealNetworks}
\begin{figure}[!th]
         		\begin{center}
                         \includegraphics[width=0.45\hsize]{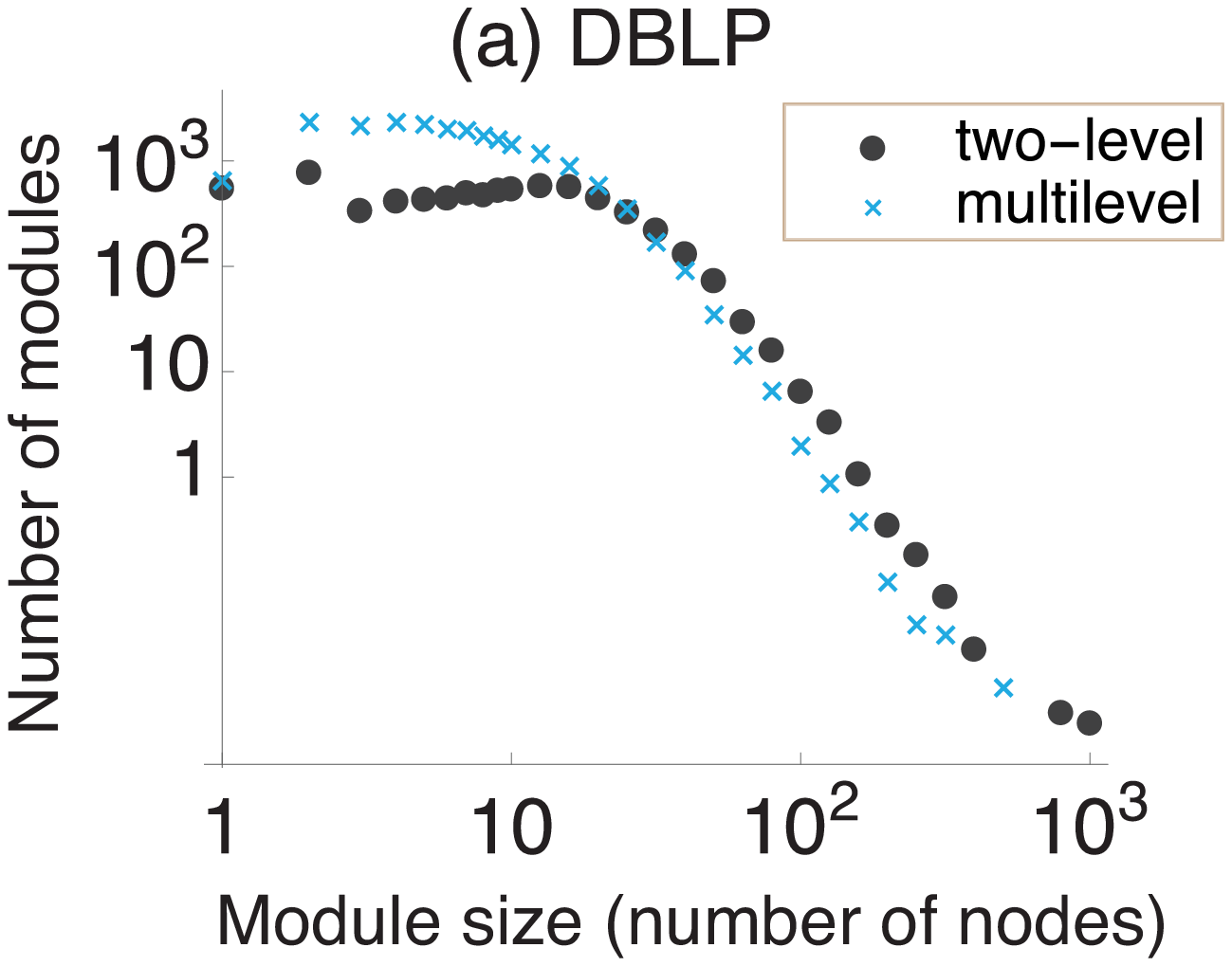}
                         \hspace{10pt}
                         \includegraphics[width=0.45\hsize]{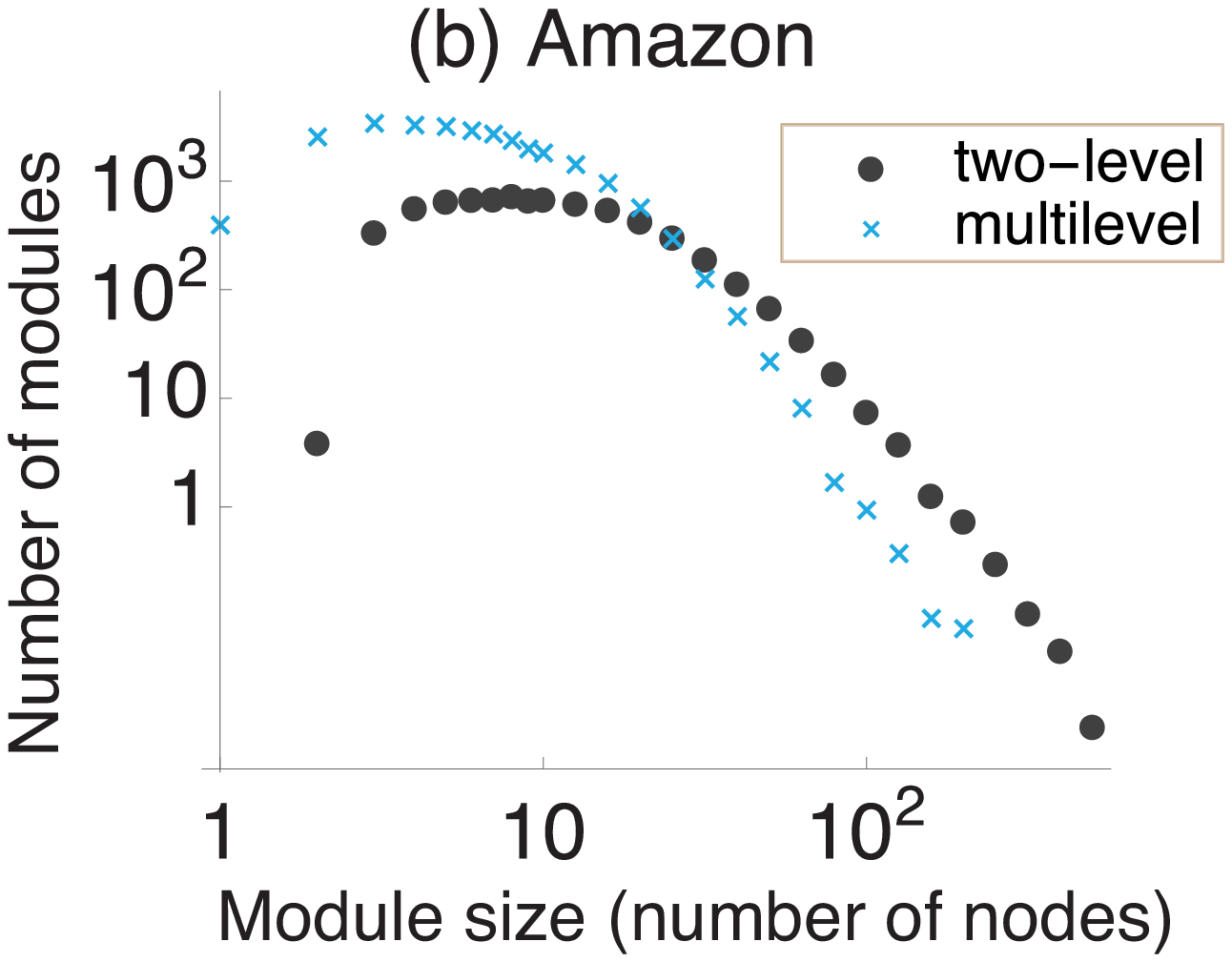}
                         \hspace{10pt}
                         \includegraphics[width=0.45\hsize]{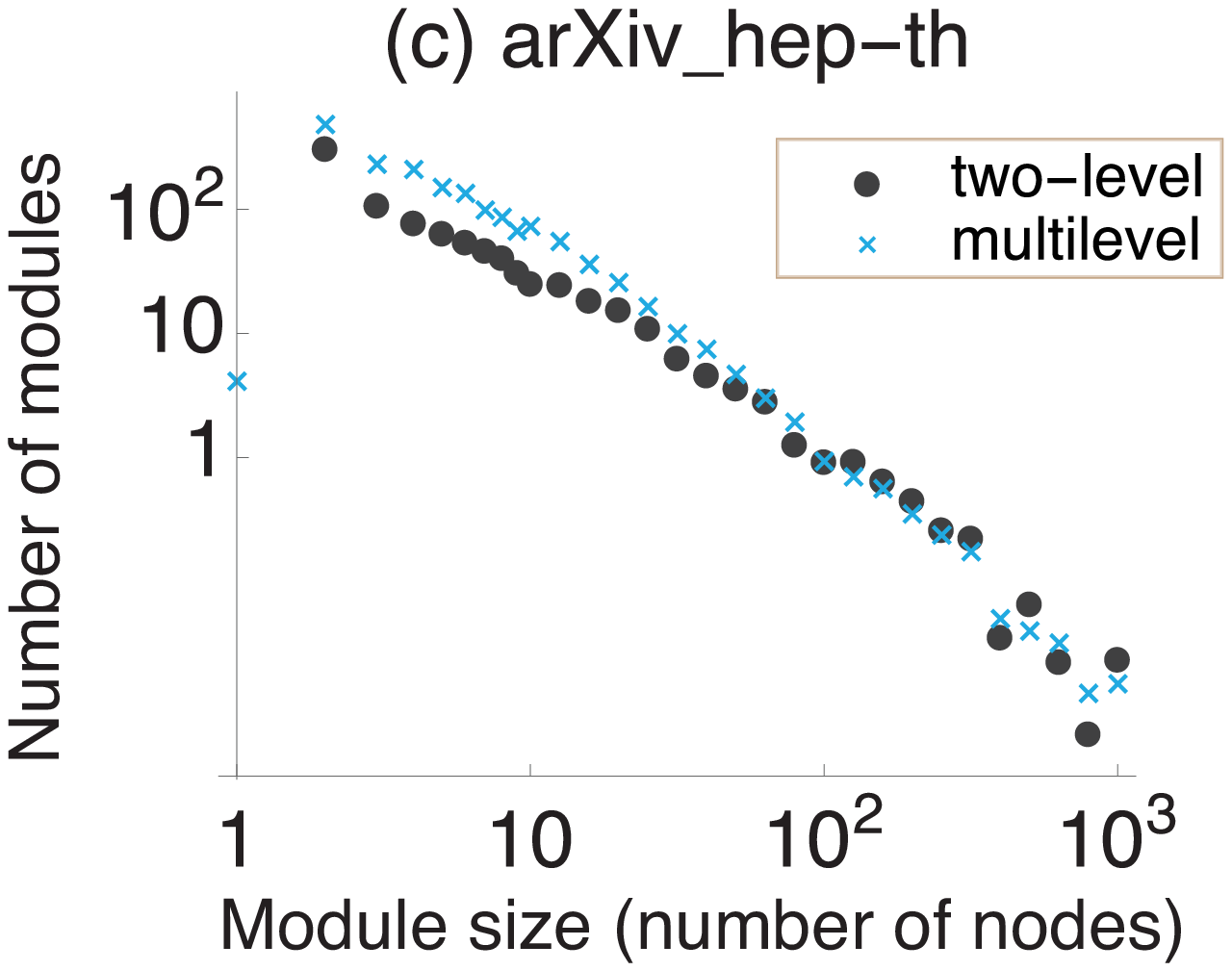}
                         \hspace{10pt}
		         \includegraphics[width=0.45\hsize]{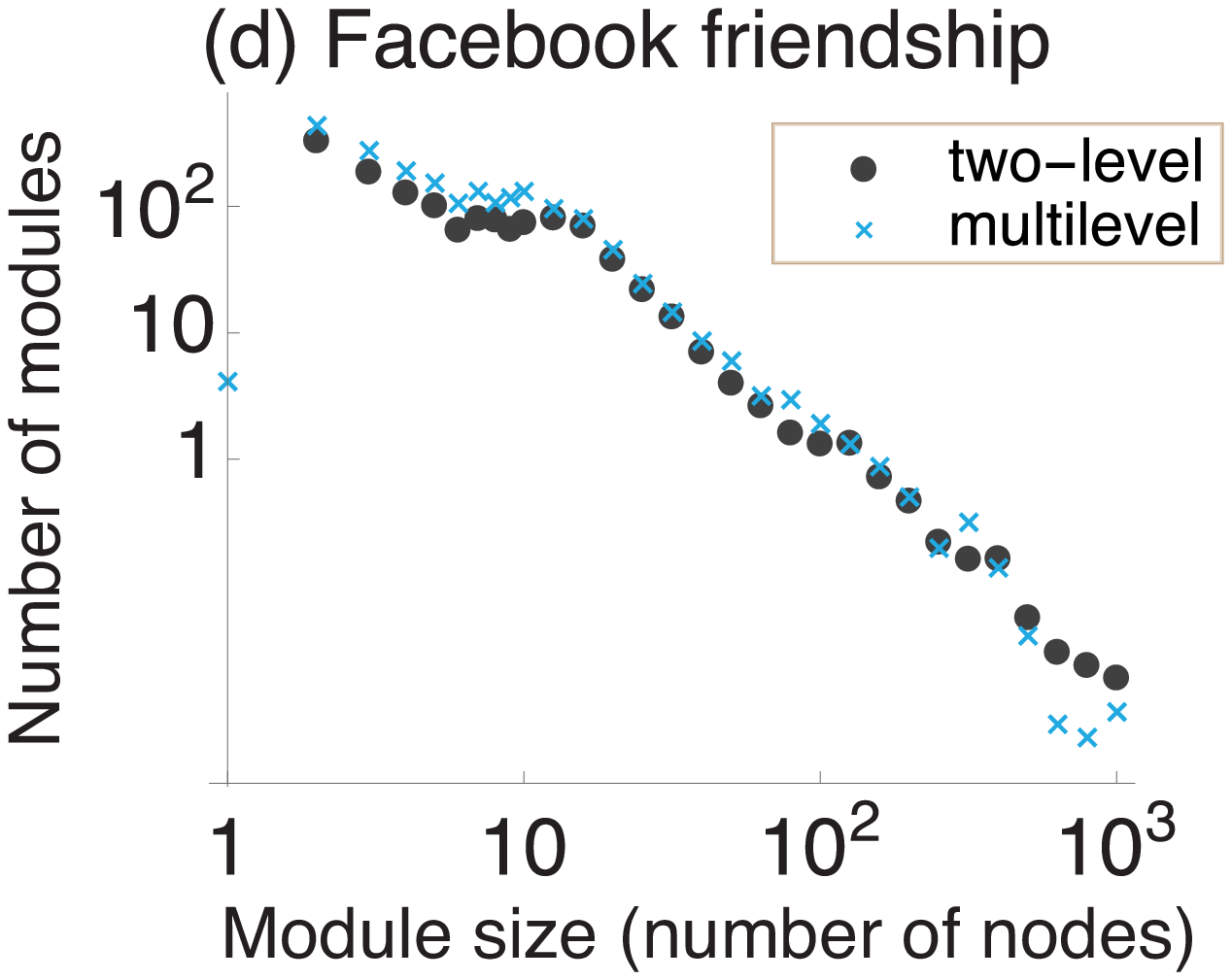}
                         \hspace{10pt}
		         \includegraphics[width=0.45\hsize]{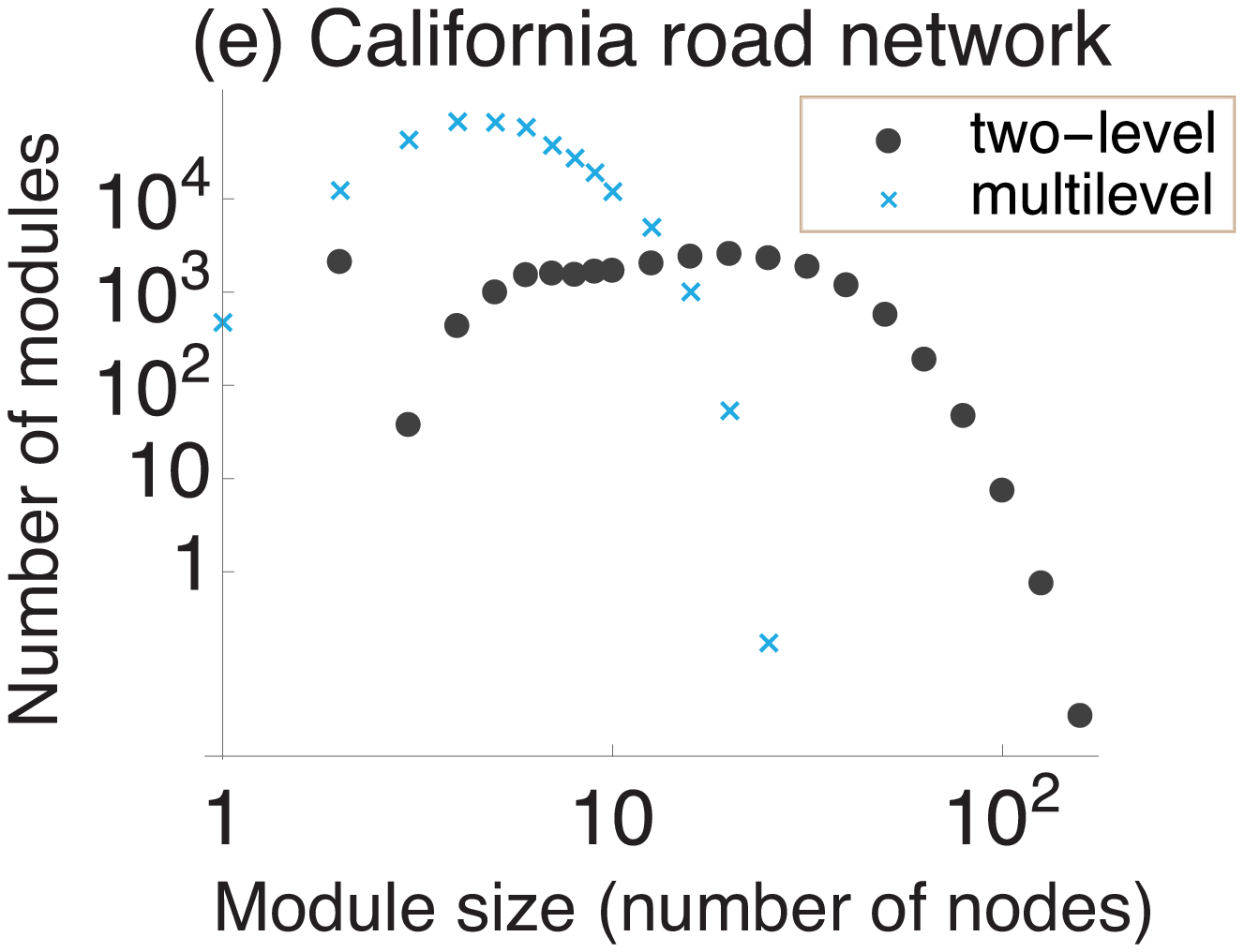}
                         \hspace{10pt}
		         \includegraphics[width=0.45\hsize]{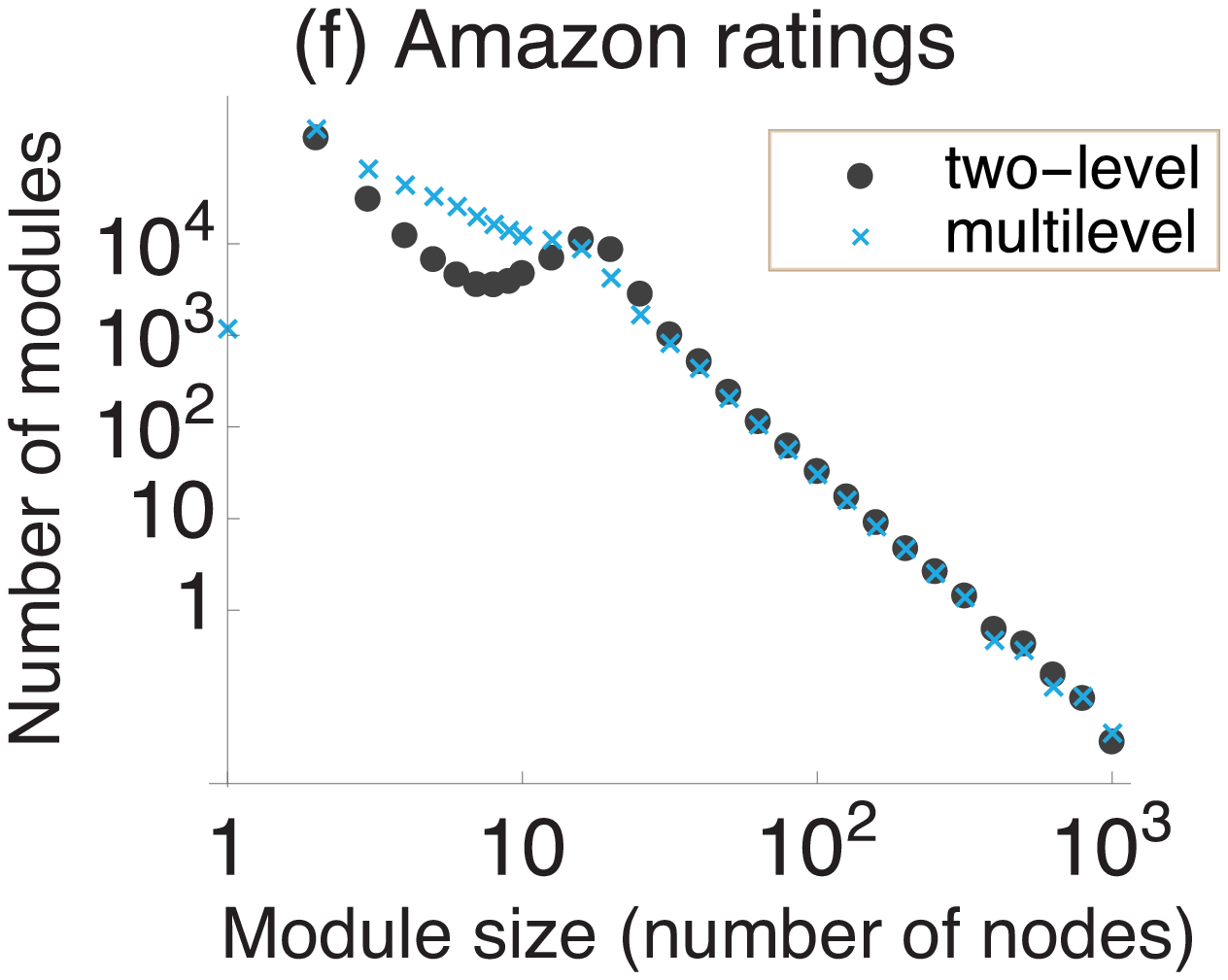}
	         	\end{center}
			\caption{(Color online) Module size distributions of 
 			(a) the collaboration network of authors of scientific papers from DBLP computer science bibliography \cite{StanfordDataset,Yang:2012:DEN:2350190.2350193}, 
			(b) the copurchasing network of products in Amazon.com \cite{StanfordDataset,Yang:2012:DEN:2350190.2350193}, 
			(c) the network of citations in arXiv's High Energy Physics -- Theory (hep-th) section \cite{b242}, 
			(d) the friendship network on Facebook \cite{b480}, 
			(e) the road network of California \cite{b336} in the U. S. A., and 
			(f) the bipartite rating network of products in Amazon.com \cite{konect:jindal2008} 
			in the log-log scale with partial-logarithmic binning. 
			The circular points represent the result of the two-level method and 
			the cross points represent the result in the finest level of the multilevel method.
			Note that the size of a module here does not indicate the number of internal links, but the number of nodes within the module. }
		         \label{RealNetworksSuppl}
         \end{figure}

In this section, we show that the relaxed resolution limit can be observed in analysis of real networks. 
Figure \ref{RealNetworksSuppl} shows the module size distributions, which we obtained by running Infomap \cite{RosvallCode} on the data distributed at \cite{StanfordDataset,konect}.
The size of each network and the total number of detected modules are listed in Table \ref{DataTable}. 
Here, size refers to the number of nodes in a module instead of the number of internal links, but the effect of the resolution limit is nevertheless clear. 
The multilevel method detects many more smaller modules \cite{MultiLevelNote}.
Note that the Amazon rating network is a bipartite network; the random walker in a bipartite network has periodic stationary states by nature, but we assume the non-periodic solution with visit rates given by Eq.~(\ref{pa}) and all derived results apply.
For some of these networks, such as the arXiv citation network and the Facebook friendship network, the effect of the resolution limit on the two-level method looks small, because these networks are dense compared to the other networks and the depth of the hierarchy in the multilevel method is shallow.

We can estimate the theoretical resolution limit by estimating the cut size in Eq.~(\ref{GreedyResolutionLimit}). The cut size is bounded below by the number of modules detected by the multilevel method and above by the number of links in the network. 
For the DBLP network, these numbers are $29,252$ and $1,049,866$, respectively, and the left-hand side of Eq.~(14) in the main text falls between these values for $l_{c} \approx 11$ \cite{MemoCommunitySize}. 

As Fig.~\ref{RealNetworksSuppl} shows, because real networks have modules of varying strength, the resolution limit does not force a clear separation between detected and undetected module sizes.
For real networks, the theoretical resolution limit is instead the point at which we can expect deviations between a two-level method and a multilevel method.


\begingroup
\begin{ruledtabular} 
\begin{table*}[!th]
\centering
\caption{Community sizes in real networks obtained with the two-level and the multilevel map equation}
\begin{tabular}{c | r | r | r | r}
	& \# of nodes & \# of links & \# of communities & \# of communities \\
	&	&	&(two-level)	& (multilevel)\\
	\hline
	DBLP & 338,029 & 1,049,866 & 16,450 & 29,252 \\ 
	Amazon (copurchase)& 334,863 & 925,872 & 15,685 & 34,802 \\ 
	Facebook & 63,731 & 1,269,502 & 2,268 & 2,819\\
	arXiv\_hep-th & 48,239 & 352,807 & 1,332 & 2,247 \\ 
	California & 1,965,206 & 2,766,607 & 82,322 & 344,485 \\
	Amazon (rating)& 3,376,972 & 5,838,041 & 350,419 & 480,810 
\end{tabular}
\label{DataTable}
\end{table*}
\end{ruledtabular}
\endgroup

\section{Conclusion}
In summary, we have revealed the inner workings of the map equation and estimated its resolution limit (Eq.~(\ref{GreedyResolutionLimit})).
While the number of links in the network determines the resolution limit of the configuration-model-based modularity,
the number of links between modules instead determines the resolution limit of the flow-based map equation.
This less restrictive dependence contributes to the performance difference between the methods. 
Even if the resolution limit in practice is many orders of magnitudes smaller than it is for modularity, for sufficiently large networks the map equation will eventually be affected, as any global two-level objective function is in practice.
We argue that the natural solution is to use the hierarchical map equation,
and exemplify both with synthetic and real networks.
We conclude that multilevel methods or Markov time sweeping should always be the first choice for simplifying large networks, but that better tools are needed for efficiently working with such structures.
Finally, we emphasize that coping with the resolution limit does not in itself imply good performance in resolving actual communities of real networks. It is still an open question as to what structures can be detected in conventional networks and when possibly higher-order information is necessary.

\section*{Acknowledgements}
The authors thank Naomichi Hatano for critical reading of the manuscript. 
T.K.\ was supported by JST, ERATO, Kawarabayashi Large Graph Project.
M.R.\ was supported by Swedish Research Council Grant No. 2012-3729.

\onecolumngrid
\appendix

\section*{APPENDIX: THE HIERARCHICAL MAP EQUATION}\label{AppendixHierarchicalMapEquation}
\subsection{General argument on partitioning}\label{AppendixHierarchicalMapEquation-General}
\begin{figure}[!t]
         		\begin{center}
		          \includegraphics[width=0.6\hsize]{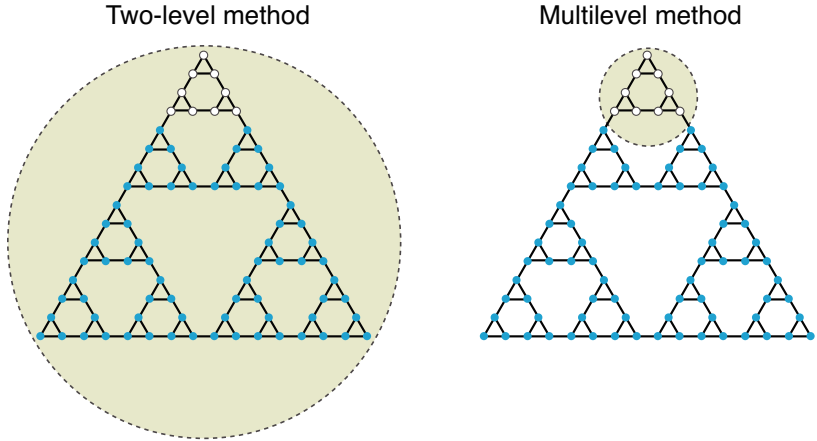}
	         	\end{center}
			\caption{(Color online) Effective network sizes (the region encircled with the dashed line) for the evaluation of partitioning of the nine nodes at the top of the Sierpinski triangle (the nodes of open circles) in the two-level method and in the finest level of the multilevel method. While the effective network size is the size of the whole network in the two-level method, in the finest level of the multilevel method, it is of the supermodule plus the links lying between the supermodule and the rest of the network. }
		         \label{MultiLevelExample}
         \end{figure}

Here we explain the details of the hierarchical map equation \cite{Rosvall2011}. 
Again, we restrict ourselves to undirected, unweighted networks. 
Analogously to the two-level method, the quality function of the multilevel method is defined by 
\begin{align}
L({\mathsf M}) &= q_{\inter} H(\mathcal{Q}) 
+ \sum^{m}_{i=1} q_{i \intra} H(\mathcal{Q}_{i}) 
+ \sum^{m}_{i=1} \sum^{m_{i}}_{j=1} q_{ij \intra} H(\mathcal{Q}_{ij}) + 
\cdots 
+ \sum_{ij\dots k} p_{ij\dots k \intra} H(\mathcal{P}_{ij\dots k}), \label{MultiLevelSuppl}
\end{align}
with 
\begin{align}
H(\mathcal{Q}) 
&= - \sum^{m}_{i=1} \frac{q_{i \inter}}{q_{\inter}} \log \frac{q_{i \inter}}{q_{\inter}}, \\
\sum^{m}_{i=1} H(\mathcal{Q}_{i}) 
&= -\sum^{m}_{i=1} \frac{q_{i \inter}}{q_{i \intra}} \log \frac{q_{i \inter}}{q_{i \intra}} 
- \sum^{m}_{i=1} \sum^{m_{i}}_{j=1} \frac{q_{ij \inter}}{q_{i \intra}} \log \frac{q_{ij \inter}}{q_{i \intra}}, \\
\sum_{ij\dots k} H(\mathcal{P}_{ij\dots k}) 
&= -\sum_{ij\dots k} \frac{q_{ij\dots k \inter}}{p_{ij\dots k \intra}} \log \frac{q_{ij\dots k \inter}}{p_{ij\dots k \intra}} 
- \sum_{ij\dots k} \sum_{\alpha \in ij\dots k} \frac{p_{\alpha}}{p_{ij\dots k \intra}} \log \frac{p_{\alpha}}{p_{ij\dots k \intra}}, 
\end{align}
where $q_{i_{1}i_{2}\dots i_{x} \inter}$ is the probability that a random walker exits from the module $i_{x}$ in the $x$th level, and 
$q_{i_{1}i_{2}\dots i_{x} \intra}$ is the probability that the random walker stays within the module $i_{x}$ and exits from it. 
As we mentioned in the main text, the probabilities of entering and exiting a module are equal for undirected networks. 
Thus, we replaced the entering probability with the exiting probability in order to simplify the notation. 
As before, $p_{\alpha}$ is the probability that the random walker visits node $\alpha$. 
Following what we did for the two-level method, we can write (\ref{MultiLevelSuppl}) as 
\begin{align}
L({\mathsf M}) 
&= q_{\inter} \log q_{\inter} + \sum_{i_{1}} q_{i_{1} \intra} \log q_{i_{1} \intra} + \sum_{i_{1}i_{2}} q_{i_{1}i_{2} \intra} \log q_{i_{1}i_{2} \intra} + \cdots 
+ \sum_{i_{1}i_{2}\dots i_{k}} q_{i_{1}i_{2}\dots i_{k} \intra} \log q_{i_{1}i_{2}\dots i_{k} \intra} 
- \sum_{\alpha} p_{\alpha} \log p_{\alpha} \nonumber\\
&\hspace{10pt} - 2 \left( \sum_{i_{1}} q_{i_{1} \inter} \log q_{i_{1} \inter} + \sum_{i_{1}i_{2}} q_{i_{1}i_{2} \inter} \log q_{i_{1}i_{2} \inter} + \cdots 
+ \sum_{i_{1}i_{2}\dots i_{k}} q_{i_{1}i_{2}\dots i_{k} \inter} \log q_{i_{1}i_{2}\dots i_{k} \inter} \right). 
\end{align}
We then consider the difference of the map equation at an update. 
The modification of the partition at a certain level does not affect the partitions in higher and lower levels. 
It only alters the description length of movements between the modules of the modified level, as well as 
the description length of movements within the modified module (it can also be regarded as movements between submodules) and exiting from it. 
Therefore, the difference of the map equation from an update in the $x$th level is 
\begin{align}
\Delta L({\mathsf M}) 
&= q^{{\mathsf B}}_{i_{1}i_{2}\dots i_{x-1} \intra} \log q^{{\mathsf B}}_{i_{1}i_{2}\dots i_{x-1} \intra} 
- q^{{\mathsf A}}_{i_{1}i_{2}\dots i_{x-1} \intra} \log q^{{\mathsf A}}_{i_{1}i_{2}\dots i_{x-1} \intra} \nonumber\\
&\hspace{10pt}-2 \left( \sum_{i^{\prime}_{x}} q^{{\mathsf B}}_{i_{1}i_{2}\dots i^{\prime}_{x} \inter} \log q^{{\mathsf B}}_{i_{1}i_{2}\dots i^{\prime}_{x} \inter} 
- \sum_{i_{x}} q^{{\mathsf A}}_{i_{1}i_{2}\dots i_{x} \inter} \log q^{{\mathsf A}}_{i_{1}i_{2}\dots i_{x} \inter}
\right) \nonumber\\ 
&\hspace{10pt}+ \left( 
\sum_{i^{\prime}_{x}} q^{{\mathsf B}}_{i_{1}i_{2}\dots i^{\prime}_{x} \intra} \log q^{{\mathsf B}}_{i_{1}i_{2}\dots i^{\prime}_{x} \intra} 
- \sum_{i_{x}} q^{{\mathsf A}}_{i_{1}i_{2}\dots i_{x} \intra} \log q^{{\mathsf A}}_{i_{1}i_{2}\dots i_{x} \intra}
\right), 
\label{HierarchicalDeltaL}
\end{align}
which is analogous to that of the two-level method, 
\begin{align}
\Delta L_{\mathrm{two-level}}({\mathsf M}) 
&= q^{{\mathsf B}}_{\inter} \log q^{{\mathsf B}}_{\inter} - q^{{\mathsf A}}_{\inter} \log q^{{\mathsf A}}_{\inter} \nonumber\\
&\hspace{10pt} -2 \left( \sum_{i^{\prime}=1}^{m^{\prime}} q^{{\mathsf B}}_{i^{\prime} \inter} \log q^{{\mathsf B}}_{i^{\prime} \inter} 
 - \sum_{i=1}^{m} q^{{\mathsf A}}_{i \inter} \log q^{{\mathsf A}}_{i \inter}
 \right) 
 + \left( 
 \sum_{i^{\prime}=1}^{m^{\prime}} p^{{\mathsf B}}_{i^{\prime} \intra} \log p^{{\mathsf B}}_{i^{\prime} \intra} 
 - \sum_{i=1}^{m} p^{{\mathsf A}}_{i \intra} \log p^{{\mathsf A}}_{i \intra}
 \right). 
\end{align}
Instead of $q_{\inter}$ in the two-level method in the above equation, we have 
\begin{align}
q_{i_{1}i_{2}\dots i_{x-1} \intra} = q_{i_{1}i_{2}\dots i_{x-1} \inter} + \sum_{i_{x}} q_{i_{1}i_{2}\dots i_{x} \inter} 
\end{align}
for the multilevel method; 
hence, other than the extra term $q_{i_{1}i_{2}\dots i_{x-1} \inter}$ in $q_{i_{1}i_{2}\dots i_{x-1} \intra}$, 
the mathematical structure of the hierarchical map equation is analogous to that of the two-level method. 
That is, the difference between these methods is the effective network size in each update process, as exemplified in Fig.~\ref{MultiLevelExample} for the Sierpinski triangle. 
The above result can also be interpreted as follows. The resolution limit of a subgraph $H$ in a graph $G$ remains the same in a subgraph $S$, where $H \subset S \subset G$, as long as the subgraph $S$ strictly includes the supermodule containing the subgraph $H$, provided that the structure of the supermodule is conserved in the graph $G$ and the subgraph $S$. 
This property is somewhat similar to a mathematical definition of a ``resolution limit-free method'' in Ref.~\cite{traag2011narrow}. 
Note, however, that the hierarchical map equation is not a resolution limit-free method in their sense, because the hierarchical structure may change for a particular choice of the subgraph $S$.

As we did for the two-level method, we can write down $\Delta L({\mathsf M})$ of the multilevel method in terms of the number of links, by using the explicit form of the stationary distribution of the random walker. 
The elements of the objective function reads 
\begin{align}
&q_{i_{1}i_{2}\dots i_{x} \inter} = \frac{l^{\mathrm{out}}_{i_{1}i_{2}\dots i_{x}}}{K}, \\
&q_{i_{1}i_{2}\dots i_{x} \intra} = \frac{2}{K}\left( l_{i_{1}i_{2}\dots i_{x}} + l^{\mathrm{out}}_{i_{1}i_{2}\dots i_{x}} \right).
\end{align}
Substituting them into Eq.~(\ref{HierarchicalDeltaL}), after some algebra, we obtain 
\begin{align}
\frac{K}{2} \Delta L({\mathsf M}) 
&= \left( l^{{\mathsf B}}_{i_{1}i_{2}\dots i_{x-1}} + l^{{\mathsf B},\mathrm{out}}_{i_{1}i_{2}\dots i_{x-1}} \right) \log \left( l^{{\mathsf B}}_{i_{1}i_{2}\dots i_{x-1}} + l^{{\mathsf B},\mathrm{out}}_{i_{1}i_{2}\dots i_{x-1}} \right) 
- \left( l^{{\mathsf A}}_{i_{1}i_{2}\dots i_{x-1}} + l^{{\mathsf A},\mathrm{out}}_{i_{1}i_{2}\dots i_{x-1}} \right) \log \left( l^{{\mathsf A}}_{i_{1}i_{2}\dots i_{x-1}} + l^{{\mathsf A},\mathrm{out}}_{i_{1}i_{2}\dots i_{x-1}} \right) \nonumber\\
&\hspace{10pt} -2 \left( l^{{\mathsf B}}_{i_{1}i_{2}\dots i_{x-1}} - l^{{\mathsf A}}_{i_{1}i_{2}\dots i_{x-1}} \right) + R_{i_{1}i_{2}\dots i_{x-1}}, 
\end{align}
where
\begin{align}
R_{i_{1}i_{2}\dots i_{x-1}} 
&= \sum_{i^{\prime}_{x}} \mathcal{L}_{i_{1}i_{2}\dots i^{\prime}_{x}}({\mathsf B}) - \sum_{i_{x}} \mathcal{L}_{i_{1}i_{2}\dots i_{x}}({\mathsf A}), \\
\mathcal{L}_{i_{1}i_{2}\dots i_{x}}({\mathsf M})
&= -l^{{\mathsf M},\mathrm{out}}_{i_{1}i_{2}\dots i_{x}} \log l^{{\mathsf M},\mathrm{out}}_{i_{1}i_{2}\dots i_{x}} 
+ \left( l^{{\mathsf M}}_{i_{1}i_{2}\dots i_{x}} + l^{{\mathsf M},\mathrm{out}}_{i_{1}i_{2}\dots i_{x}} \right) \log \left( l^{{\mathsf M}}_{i_{1}i_{2}\dots i_{x}} + l^{{\mathsf M},\mathrm{out}}_{i_{1}i_{2}\dots i_{x}} \right). 
\end{align}
Denoting $C_{i_{1}i_{2}\dots i_{x}} \equiv l_{i_{1}i_{2}\dots i_{x-1}} + l^{\mathrm{out}}_{i_{1}i_{2}\dots i_{x-1}}$ and 
letting $\delta$ be the difference of $C_{i_{1}i_{2}\dots i_{x}}$ under an update, we have
\begin{align}
\frac{K}{2} \Delta L({\mathsf M}) 
&\simeq - \delta \left( 2 + \log\left( \mathrm{e} \, C_{i_{1}i_{2}\dots i_{x}} \right) \right) + R_{i_{1}i_{2}\dots i_{x-1}}, \label{MultiLevelgeneralDeltaL}
\end{align}
where we assumed $ \delta \ll l_{i_{i_{1}i_{2}\dots x-1}}$.
Equation (\ref{MultiLevelgeneralDeltaL}) corresponds to Eq.~(\ref{generalDeltaL}) in the main text. With this correspondence, the same argument holds for the multilevel method as for the two-level method. 
Since Eq.~(\ref{MultiLevelgeneralDeltaL}) depends only on the structure of a subnetwork instead of the full network, the multilevel method has higher resolution than the two-level method, provided that the network has a nested multilevel modular structure.

\subsection{Generation of supermodules} \label{AppendixHierarchicalMapEquation-SupermoduleGeneration}

We showed that the existence of the nested structure of modules in the multilevel method of the map equation enable us to resolve smaller modules than the two-level method. 
When the multilevel method generates only two levels, however, the multilevel method is equivalent to the two-level method, and the resolution limit would be kept the same. 
As we did for the update of partition in the two-level method in the main text, we can evaluate the generation of a higher level by comparing the partitions with and without a supermoudule. 
For simplicity, we compare a two-level partition and a three-level partition. 
The objective functions of the map equation with two-level structure $L({\mathsf M}_{2})$ and three-level structure $L({\mathsf M}_{3})$ read 
\begin{align}
L({\mathsf M}_{2}) &= q^{(2)}_{\inter} \log q^{(2)}_{\inter} - 2 \sum_{i^{\prime}=1}^{m^{\prime}} q^{(2)}_{i^{\prime} \inter} \log q^{(2)}_{i^{\prime} \inter} + \sum_{i^{\prime}=1}^{m^{\prime}} p^{(2)}_{i^{\prime} \intra} \log p^{(2)}_{i^{\prime} \intra} - \sum_{\alpha} p_{\alpha} \log p_{\alpha}, \label{General2level}\\
L({\mathsf M}_{3}) &= q^{(3)}_{\inter} \log q^{(3)}_{\inter} - 2 \sum_{i=1}^{m} q^{(3)}_{i \inter} \log q^{(3)}_{i \inter} + \sum_{i=1}^{m} p^{(3)}_{i \intra} \log p^{(3)}_{i \intra} \nonumber\\ 
&\hspace{10pt} - 2 \sum_{i=1}^{m} \sum_{j=1}^{m_{i}} q^{(3)}_{ij \inter} \log q^{(3)}_{ij \inter} + \sum_{i=1}^{m} \sum_{j=1}^{m_{i}} p^{(3)}_{ij \intra} \log p^{(3)}_{ij \intra}
- \sum_{\alpha} p_{\alpha} \log p_{\alpha}. \label{General3level}
\end{align}

\begin{figure}[!t]
         		\begin{center}
		          \includegraphics[width=0.9\hsize]{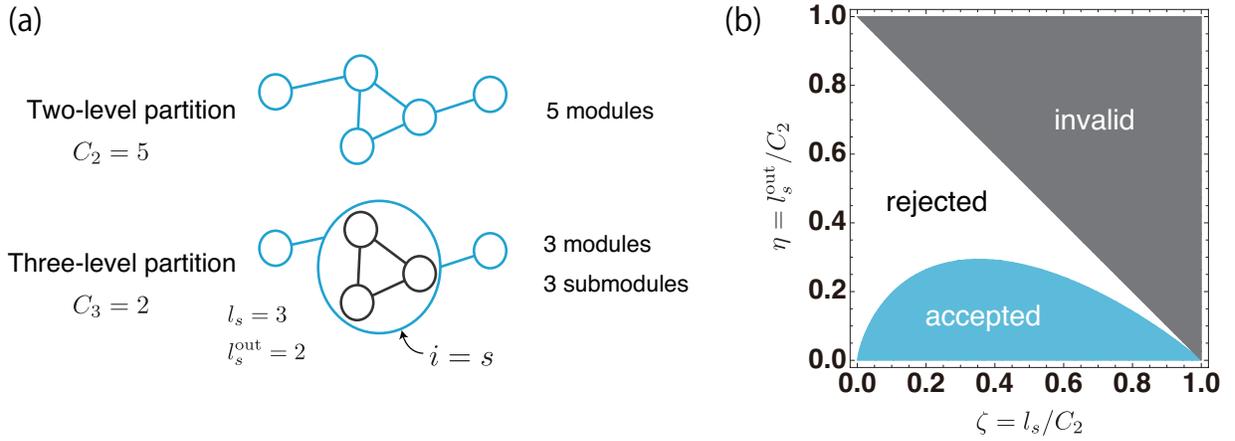}
	         	\end{center}
			\caption{(Color online)
			(a) An example of a hierarchical community structure in a network. 
			(b) Phase diagram of the accepted and rejected regions for the generation of a higher level. The shaded region (\ref{C2bound}) of the upper-right corner is invalid.
			}
		         \label{GraphHierarchyPhaseDiagram}
         \end{figure}

We consider the difference of the description length given by the map equation when we introduce a supermodule $s$ in addition to the two-level partition, forming a three-level partition. 
In the following, we denote the label of a module in the three-level structure by $i$ and the label of a submodule in module $s$ by $j$ [see Fig.~\ref{GraphHierarchyPhaseDiagram}(a)]. 
Based on Eqs.~(\ref{General2level}) and (\ref{General3level}), we have 
\begin{align}
L({\mathsf M}_{3}) - L({\mathsf M}_{2}) 
&= q^{(3)}_{\inter} \log q^{(3)}_{\inter} - 2 q^{(3)}_{s \inter} \log q^{(3)}_{s \inter} + p^{(3)}_{s \intra} \log p^{(3)}_{s \intra} 
- 2 \sum_{j=1}^{m_{s}} q^{(3)}_{sj \inter} \log q^{(3)}_{sj \inter} + \sum_{j=1}^{m_{s}} p^{(3)}_{sj \intra} \log p^{(3)}_{sj \intra} \nonumber\\
&\hspace{30pt} - \left( q^{(2)}_{\inter} \log q^{(2)}_{\inter} - 2 \sum_{j=1}^{m_{s}} q^{(2)}_{j \inter} \log q^{(2)}_{j \inter} + \sum_{j=1}^{m_{s}} p^{(2)}_{j \intra} \log p^{(2)}_{j \intra} \right) \nonumber\\
&= q^{(3)}_{\inter} \log q^{(3)}_{\inter} - q^{(2)}_{\inter} \log q^{(2)}_{\inter} - 2 q^{(3)}_{s \inter} \log q^{(3)}_{s \inter} + p^{(3)}_{s \intra} \log p^{(3)}_{s \intra}, \label{Delta23-1}
\end{align}
where 
\begin{align}
q^{(2)}_{\inter} = \sum_{i=1}^{m} q_{i \inter} - q_{s \inter} + \sum_{j=1}^{m_{s}} q_{s j \inter}, \hspace{30pt}
q^{(3)}_{\inter} = \sum_{i=1}^{m} q_{i \inter}. 
\end{align}
Note here that $q^{(3)}_{sj \inter} = q^{(2)}_{j \inter}$ and $p^{(3)}_{sj \intra} = p^{(2)}_{j \intra}$. 
When Eq.~(\ref{Delta23-1}) is negative, the construction of an additional level is accepted, i.e., the three-level description gives a shorter code length.

For undirected networks, Eq.~(\ref{Delta23-1}) can be written in terms of the number of links as follows: 
\begin{align}
&L({\mathsf M}_{3}) - L({\mathsf M}_{2}) \nonumber\\
&= \frac{\sum_{i} l^{\mathrm{out}}_{i}}{K} \log \frac{\sum_{i} l^{\mathrm{out}}_{i}}{K} 
- \left( \frac{\sum_{i} l^{\mathrm{out}}_{i} + 2l_{s} }{K} \right) \log 
\left( \frac{\sum_{i} l^{\mathrm{out}}_{i} + 2l_{s} }{K} \right) 
- 2 \frac{l^{\mathrm{out}}_{s}}{K} \log \frac{l^{\mathrm{out}}_{s}}{K} 
+ \left(2 \frac{l_{s} + l^{\mathrm{out}}_{s}}{K} \right) \log \left(2 \frac{l_{s} + l^{\mathrm{out}}_{s}}{K} \right) \nonumber\\
&= \frac{2}{K} \biggl[ 
C_{3} \log C_{3} - \left(C_{3} + l_{s}\right) \log \left( C_{3} + l_{s} \right)
- l^{\mathrm{out}}_{s} \log l^{\mathrm{out}}_{s} 
+ ( l^{\mathrm{out}}_{s} + l_{s} ) \log \left( l^{\mathrm{out}}_{s} + l_{s} \right) + l^{\mathrm{out}}_{s}
\biggr]
, \label{Delta23-2}
\end{align}
where 
$l_{s}$, $l^{\mathrm{out}}_{i}$, and $C_{3} = \sum_{i} l^{\mathrm{out}}_{i}/2$ are variables of the coarsest level in the three-level partition, which are 
the number of links within the module $s$, 
the number of links between nodes within module $i$ and nodes outside the module, and 
the cut size of the network, respectively. 

As in the main text, we introduce 
\begin{align}
\mathcal{L}(x,y) &= -x \log x + (x+y) \log (x+y) \hspace{10pt}(x>0\text{ and }y>0), \label{LocalL}
\end{align}
which has the following properties: 
\begin{align}
&\frac{\partial \mathcal{L}(x,y)}{\partial x} = \log \left( 1 + \frac{y}{x} \right) > 0, \label{Monotonicity}\\
&\mathcal{L}(x-y, y) = -\mathcal{L}(x, -y) \hspace{10pt}\text{for }x \ge y, \label{Symmetry}\\
&\mathcal{L}(a x, a y) = ay \log a + a \mathcal{L}(x, y) \hspace{10pt}\text{for }a>0. \label{ScalingProperty}
\end{align}
Using $\mathcal{L}(x,y)$, we can recast Eq.~(\ref{Delta23-2}) as 
\begin{align}
\frac{K}{2} \left( L({\mathsf M}_{3}) - L({\mathsf M}_{2}) \right) = \mathcal{L}(l^{\mathrm{out}}_{s}, l_{s}) - \mathcal{L}(C_{3}, l_{s}) + l^{\mathrm{out}}_{s}. \label{Delta23-3}
\end{align}
Therefore, the transition to the three-level structure occurs when the following condition is satisfied: 
\begin{align}
l^{\mathrm{out}}_{s} < \mathcal{L}(C_{3}, l_{s}) - \mathcal{L}(l^{\mathrm{out}}_{s}, l_{s}). \label{Transition3}
\end{align}
The right-hand-side of (\ref{Transition3}) is always greater than or equal to zero, since $\mathcal{L}(x,y)$ has monotonicity (\ref{Monotonicity}) and $C_{3} \ge l^{\mathrm{out}}_{s}$.
If the partition at the coarsest level is a bisection, it would give $C_{3} = l^{\mathrm{out}}_{s}$, which never satisfies Eq.~(\ref{Transition3}). 

In terms of the cut size in the two-level partition $C_{2} = C_{3} + l_{s}$, Eq.~(\ref{Transition3}) can also be written as 
\begin{align}
l^{\mathrm{out}}_{s} < \mathcal{L}(C_{2} - l_{s}, l_{s}) - \mathcal{L}(l^{\mathrm{out}}_{s}, l_{s}), \label{Transition2-1} 
\end{align}
or using (\ref{Symmetry}) and (\ref{ScalingProperty}), we have 
\begin{align}
0 > \eta + \mathcal{L}(1, - \zeta) +  \mathcal{L}(\eta, \zeta), \label{Transition2-3} 
\end{align}
where $\zeta = l_{s} / C_{2}$ and $\eta = l^{\mathrm{out}}_{s} / C_{2}$.
The phase diagram of the generation of a higher level determined by Eq.~(\ref{Transition2-3}) is shown in Fig.~\ref{GraphHierarchyPhaseDiagram}(b). 
Note that the cut size $C_{2}$ is bounded below by $l_{s} + l^{\mathrm{out}}_{s} \le C_{2}$; i.e., 
\begin{align}
\zeta + \eta \le 1. \label{C2bound}
\end{align}
The boundary in (\ref{Transition2-3}) never intersects with the boundary of the invalid region (\ref{C2bound}). 
This observation can be confirmed from Eq.~(\ref{Transition3}), which never satisfies the inequality at the boundary of the invalid region $l^{\mathrm{out}}_{s} = C_{3} = C_{2}-l_{s}$. 

Notice that the analysis here is not the whole story of the optimization of the hierarchical partitioning. 
The path for optimization can be very complex, because the generation of the deeper hierarchy and the update of partition in each level occur simultaneously. 
Roughly speaking, while dense networks tend to have large modules with shallow hierarchies, the hierarchies of sparse networks tend to be deep, i.e. the multilevel method has very high resolution. 
This behavior can be confirmed in a synthetic graph (Sec.~\ref{AppendixHierarchicalMapEquation-HierarchicalRing}), as well as in real networks (Sec.~\ref{RealNetworks}).
As we exemplify in the next section, however, we can use the result here in order to check whether an obtained partition is truly the optimal solution, or can be improved at least by adding another level.

\subsection{Multi-level solution of a ring of cliques} \label{AppendixHierarchicalMapEquation-HierarchicalRing}
In the previous section, we observed in Fig.~\ref{GraphHierarchyPhaseDiagram}(b) that the generation of higher levels will not be accepted when the network is only weakly modular. 
In this section, we exemplify with a ring of cliques that a very modular network indeed has a nested multilevel structure and that the modules of the finest levels are always resolved. 

We set the number of cliques equal to $m$ and refer to the number of links within a module as $l$, as in the main text. 
We also denote the number of modules in each level as $m_{g}$ ($g=1, 2, \dots, d-1$). 
In the case of a ring of cliques, the number of links connected to nodes outside of a module is two at any level. 
Moreover, due to the symmetry of the graph, the sizes of modules for the same level must be equal. 
With the $d$-level map equation, we have the following objective function: 
\begin{align}
K L({\mathsf M}) 
&= 2m_{1} \log (2m_{1}) + m_{1}(2 + 2 m_{2}) \log (2 + 2 m_{2}) + m_{1}m_{2}(2 + 2 m_{3}) \log (2 + 2 m_{3}) + \cdots \nonumber\\
&\hspace{10pt}+ \prod_{g=1}^{d-1} m_{g}\bigl( 2 + (2l + 2) \bigr) \log (2 + (2l + 2)) 
- 2 \left( 2m_{1} + 2m_{1}m_{2} + \cdots + 2\prod_{g=1}^{d-1} m_{g} \right) - \sum_{\alpha} k_{a} \log k_{\alpha}, \\
\frac{K}{2} L({\mathsf M}) 
&= m_{1} \log m_{1} 
+ \sum_{k=2}^{d-1} \prod_{g=1}^{k-1} m_{g} (1+m_{k}) \log (1+m_{k}) 
+ \prod_{g=1}^{d-1} m_{g} \biggl( (l+1) + (l+2) \log (l+2) \biggr) - \frac{1}{2}\sum_{\alpha} k_{a} \log k_{\alpha}. \label{ringL}
\end{align}
Also, since the number of finest modules $m$ is fixed, we have the constraint 
\begin{align}
m = \prod_{g=1}^{d-1} m_{g}. \label{NumberOfModules}
\end{align}
Hence, the optimal solution is obtained by minimizing (\ref{ringL}) subject to (\ref{NumberOfModules}). 
Although the number of modules of each level $m_{g}$ is an integer, if we approximate it as a continuous variable, the present problem becomes equivalent to solving the following Lagrange multiplier: 
\begin{align}
&\delta \biggl[ m_{1} \log m_{1} 
+ \sum_{k=2}^{d-1} \prod_{g=1}^{k-1} m_{g} (1+m_{k}) \log (1+m_{k}) + \prod_{g=1}^{d-1} m_{g} \biggl( (l+1) + (l+2) \log (l+2) \biggr) - \lambda \left(\prod_{g=1}^{d-1} m_{g} - m \right) \biggr] \nonumber\\
&= \delta \biggl[ m_{1} \log m_{1} 
+ \sum_{k=2}^{d-1} \prod_{g=1}^{k-1} m_{g} (1+m_{k}) \log (1+m_{k}) + \prod_{g=1}^{d-1} m_{g} \Xi(l, \lambda) 
+ \lambda m \biggr] = 0, 
\end{align}
where we set 
\begin{align}
\Xi(l, \lambda) \equiv (l+1) + (l+2) \log (l+2) - \lambda.
\end{align}
After some straightforward calculations, we obtain the value of $m_{g}$ as a function of $m_{g+1}$ as follows: 
\begin{align}
&m_{1} = \frac{\mathrm{e}^{m_{2}-1}}{1 + m_{2}}, \nonumber\\
&m_{g} = \frac{\mathrm{e}^{m_{g+1}-1}}{1 + m_{g+1}} - 1 \hspace{30pt} \text{for } 2 \le g \le d-2, \nonumber\\
&m_{d-1} = \frac{2^{\Xi}}{\mathrm{e}} - 1. 
\end{align}
The value of $\lambda$ in $\Xi(l, \lambda)$ is determined by the constraint of $m$, the total number of modules. 
Hence, the number of modules at the deepest level in a supermodule $m_{d-1}$ is determined by $m$, $d$, and $l$. 
In the actual multilevel method, the number of hierarchal levels $d$ is adjusted so that the average code length is minimized. 

	\begin{figure}[t]
         	\begin{center}
		\includegraphics[width=0.27\hsize]{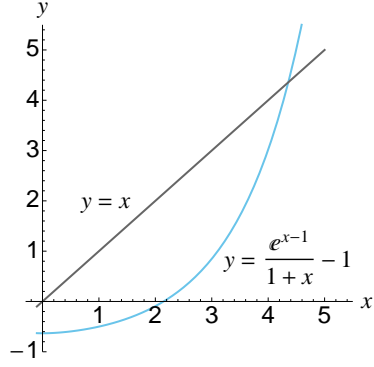}
	         \end{center}
		\caption{(Color online) The plot $y = f(x)$ in Eq.~(\ref{fx}).}
		\label{mgCondition}
         \end{figure}       

We now estimate the values of $m_{d-1}$. 
Defining 
\begin{align}
f(x) = \frac{\mathrm{e}^{x-1}}{1 + x} - 1, \label{fx}
\end{align}
we have 
$f(3) = 0.85$, 
$f(4) = 3.02$, and 
$f(5) = 8.10$ 
[see Fig.~\ref{mgCondition} for the shape of $f(x)$].
First, we readily see that $m_{d-1} \le 3$ is not the choice for a large ring, because the number of modules in a higher level decreases in such cases, which restricts the value of $m$ in (\ref{NumberOfModules}). 
If $m_{d-1} \ge 5$ and we assume that there are more than two levels, $m_{1}$ would be very large. 
However, this is unlikely the optimal, because in the phase diagram of Fig.~\ref{GraphHierarchyPhaseDiagram}(b), when $l_{s}^{\mathrm{out}} = 2$ and $m_{1}$ is large enough, there must be a supermodule of a higher level which makes the total description length shorter.  
Therefore, the number of modules inside of a supermodule tends to be close to four in the ring of cliques. 
Accordingly, every clique is always detectable with the multilevel method no matter how large the size of the ring is.

\bibliographystyle{apsrev}

\bibliography{Infomap_Resolution}

\begin{thebibliography}{37}
\bibitem[{Mul()}]{MultiLevelNote}
\bibinfo{howpublished}{The multilevel solution is not constructed as a simple
  decomposition of the two-level solution, but by an attempt to find a global
  optimal modular description. Therefore, the multilevel method does not
  trivially detect small modules more than the two-level method does.}
\bibitem[{Mem()}]{MemoCommunitySize}
\bibinfo{howpublished}{Note that the number of nodes $n$ within a module is
  bounded by the number of links $l$ within the module by $n \leq l+1$. The
  equality holds for a module of a tree. Therefore, the results of the
  two-level method and the multilevel method should deviate from each other by
  about $n \simeq l$.}
\end{thebibliography}



\end{document}